\documentclass[preprintnumbers,amsmath,amssymb,showpacs]{revtex4}

\usepackage{graphicx}
\usepackage{dcolumn}
\usepackage{bm}
\usepackage{epstopdf}%
\usepackage{caption}%

\begin{document}

\title{Tighter monogamy and polygamy relations of multiparty quantum entanglement}

\author{Limin Gao}
\author{Fengli Yan}
\email{flyan@hebtu.edu.cn}
\affiliation {College of Physics Science and Information Engineering, Hebei
Normal University, Shijiazhuang 050024, China}
\author{Ting Gao}
\email{gaoting@hebtu.edu.cn}
\affiliation {College of Mathematics and Information Science, Hebei
Normal University, Shijiazhuang 050024, China}

\date{\today}

\begin{abstract}
We investigate the tight monogamy and polygamy relations of multiparty entanglement for arbitrary quantum states. By using the power of the bipartite measure of entanglement, we establish a class of tight monogamy relations of multiparty entanglement with larger lower bounds than the existing monogamy relations. We also give a class of tight polygamy relations of multiparty entanglement with smaller upper bounds than the existing polygamy relations, by using the  power of the entanglement of assistance. It is shown that these new monogamy and polygamy relations are tighter than the former results.
\end{abstract}


\maketitle

\section{INTRODUCTION}

An important property of entanglement is the monogamy of entanglement (MOE) [1, 2]. It says that entanglement cannot be freely shared unconditionally among the multipartite quantum systems. For example, for three two-level quantum systems, denoted by \emph{A}, \emph{B} and \emph{C}, if \emph{A} and \emph{B} are in a maximally entangled state $|\Psi^-\rangle=(|01\rangle-|10\rangle)/\sqrt{2}$, then \emph{A }cannot be entangled to \emph{C}. This indicates that it should obey some trade-off on the amount of entanglement between the pairs \emph{AB} and \emph{AC}. The first mathematical characterization of MOE was expressed as a form of inequality for three-qubit state in terms of squared concurrence, which was generalized to arbitrary multiqubit systems by Osborne and Verstraete [3]. Later, the same monogamy inequality was also generalized to other entanglement measures [4-10]. Monogamy relations will help us to a further understanding of the distribution of entanglement in multipartite systems. Moreover, it also plays an important role in quantum information theory [11], condensed-matter physics [12] and even black-hole physics [13].

On the other hand the assisted entanglement, which is a dual amount to bipartite entanglement measures, is also shown to have a polygamous of entanglement (POE). POE can be considered as another kind of entanglement constraints in multiparty quantum systems. A polygamy inequality was first established for arbitrary multiqubit systems by using the squared concurrence of assistance [14-16]. Later, it was shown that the same polygamy inequality was also generalized in terms of various assisted entanglements [17-19]. Recently, a class of tight monogamy relations and polygamy relations were derived in multiparty quantum systems [20-28]. In this paper, we establish new classes of tight monogamy and polygamy relations of multiparty entanglement for arbitrary quantum states, based on the power of the bipartite measure of entanglement and the entanglement of assistance. We show that these new monogamy and polygamy relations are tighter than the results in [20-27].

\section{tighter monogamy relations of multiparty quantum entanglement}

We denote the state of a multipartite system with a finite dimensional Hilbert space $\mathcal{H}_{A}\otimes \mathcal{H}_{B_1}\otimes\cdots\otimes\ \mathcal{H}_{B_{N-1}}$ by $\rho_{A|B_{1}\cdots B_{N-1}}$. Given any a bipartite entanglement measure $E$ of the quantum states, $E$ is said to be monogamous if the following inequality is satisfied
\begin{equation}\label{}
E^{\alpha_{c}}(\rho_{A|B_{1}\cdots B_{N-1}})\geq \sum\limits_{i=1}^{N-1} E^{\alpha_{c}}(\rho_{A|B_{i}}),
\end{equation}
where $\rho_{A|B_{i}}=\text{tr}_{B_{1}\cdots B_{i-1}B_{i+1}\cdots B_{N-1}}(\rho_{A|B_{1}\cdots B_{N-1}})$, $\alpha_{c}$ is the infimum exponent for $E^{\alpha_{c}}$ to be monogamous. In order to investigate the monogamy relations of multiparty quantum entanglement, we need the following lemmas.

\textbf{Lemma 1.} For $x\geq m\geq1$ and $\mu\geq1$, then
\begin{equation}\label{}
(1+x)^{\mu}\geq x^{\mu}+(m+1)^{\mu}-m^{\mu}.
\end{equation}

Proof. Let $f(\mu,x)=(1+x)^{\mu}-{x^{\mu}}$. Then, $\frac{\partial f}{\partial x}= \mu[(1+x)^{\mu-1}-x^{\mu-1}]$. When $x\geq m\geq1$ and $\mu\geq1$, it is obviously that $(1+x)^{\mu-1}\geq x^{\mu-1}$. Thus, $\frac{\partial f}{\partial x}\geq0$, $f(\mu,x)$ is an increasing function of $x$, i.e. $ f(\mu,x)\geq f(\mu,m)=(m+1)^{\mu}-m^{\mu}$. Thus we have $(1+x)^{\mu}\geq x^{\mu}+(m+1)^{\mu}-m^{\mu}$.

\textbf{Lemma 2.} For $a_{1}\geq a_{2}\geq \cdots \geq a_{n}\geq0$ and $\mu\geq1$, then
\begin{equation}\label{}
(a_{1}+a_{2}+\cdots+a_{n})^{\mu} \geq a_{1}^{\mu}+(2^{\mu}-1)a_{2}^{\mu}+\cdots+[n^{\mu}-(n-1)^{\mu}]a_{n}^{\mu}.
\end{equation}

Proof. For the case that $n=1$, the inequality (3) is trivial. Now we assume $n=m$ the inequality (3) holds with $m>1$ and consider the case that $n=m+1$. If $a_{m+1}=0$, the inequality is trivial. Otherwise, let $\tau=\frac{a_{1}+a_{2}+\cdots+a_{m}}{a_{m+1}}$, since $a_{1}\geq a_{2}\geq \cdots \geq a_{m+1}>0$, then $\tau\geq m$,

\begin{equation}\label{}
\begin{aligned}
(a_{1}+a_{2}+\cdots+a_{m}+a_{m+1})^{\mu}= & a_{m+1}^{\mu}(1+\frac{a_{1}+a_{2}+\cdots+a_{m}}{a_{m+1}})^{\mu}\\
= & a_{m+1}^{\mu}(1+\tau)^{\mu}\\
\geq & a_{m+1}^{\mu}\left[\tau^{\mu}+(m+1)^{\mu}-m^{\mu}\right]\\
= & (a_{1}+a_{2}+\cdots+a_{m})^{\mu}+[(m+1)^{\mu}-m^{\mu}]a_{m+1}^{\mu},
\end{aligned}
\end{equation}
where the inequality is due to the inequality (2).

The induction hypothesis yields
\begin{equation}\label{}
\begin{aligned}
(a_{1}+a_{2}+\cdots+a_{m})^{\mu} \geq a_{1}^{\mu}+(2^{\mu}-1)a_{2}^{\mu}+\cdots+[m^{\mu}-(m-1)^{\mu}]a_{m}^{\mu}.
\end{aligned}
\end{equation}
Combining inequalities (4) and (5), we have
\begin{equation}\label{}
\begin{aligned}
(a_{1}+a_{2}+\cdots+a_{m+1})^{\mu} \geq a_{1}^{\mu}+(2^{\mu}-1)a_{2}^{\mu}+\cdots+[(m+1)^{\mu}-m^{\mu}]a_{m+1}^{\mu}.
\end{aligned}
\end{equation}
It implies that the inequality (3) holds for the case that $n=m+1$, which completes the proof of Lemma 2.

\textbf{Theorem 1.} For an $N$-party state $\rho_{A|B_{1}\cdots B_{N-1}}\in \mathcal{H}_{A}\otimes \mathcal{H}_{B_{1}}\otimes\cdots\otimes \mathcal{H}_{B_{N-1}}$, postulate $E^{\alpha_{c}}$ is a monogamous entanglement measure of the quantum states. If $E(\rho_{A|B_{i}})\geq E(\rho_{A|B_{i+1}})$ for $i=1,2,\cdots,N-2$, $N\geq 3$ then
\begin{equation}\label{}
 \begin{aligned}
 E^{\eta}(\rho_{A|B_{1}\cdots B_{N-1}}) \geq
 E^{\eta}(\rho_{A|B_{1}})+(2^{t}-1)E^{\eta}(\rho_{A|B_{2}})+\cdots
 +[(N-1)^{t}-(N-2)^{t}]E^{\eta}(\rho_{A|B_{N-1}})
\end{aligned}
\end{equation}
for $\eta\geq\alpha_{c}$ and $t=\frac{\eta}{\alpha_{c}}$.

Proof. Without loss of generality, the condition  $E(\rho_{A|B_{i}})\geq E(\rho_{A|B_{i+1}})$ can be always satisfied by relabeling the subsystems.
From the inequality (1), one has
\begin{equation}\label{}
 \begin{aligned}
 E^{\eta}(\rho_{A|B_{1}\cdots B_{N-1}}) \geq
[E^{\alpha_{c}}(\rho_{A|B_1})+E^{\alpha_{c}}(\rho_{A|B_2})+\cdots+E^{\alpha_{c}}(\rho_{A|B_{N-1}})]^{t}.
\end{aligned}
\end{equation}

 If $E(\rho_{A|B_{i}})\geq E(\rho_{A|B_{i+1}})$ for $i=1,2,\cdots,N-2$, according to Lemma 2, one gets
\begin{equation}\label{}
 \begin{aligned}
 E^{\eta}(\rho_{A|B_{1}\cdots B_{N-1}}) \geq
 E^{\eta}(\rho_{A|B_{1}})+(2^{t}-1)E^{\eta}(\rho_{A|B_{2}})+\cdots
 +[(N-1)^{t}-(N-2)^{t}]E^{\eta}(\rho_{A|B_{N-1}}).
\end{aligned}
\end{equation}

\textbf{Remark 1}. It is easy to verify that Theorem 1 is generally tighter than the monogamy relations in terms of the Hamming weight [27].

For later use we prove the following lemma.

\textbf{Lemma 3.} For $0\leq x\leq \frac{1}{k}, k\geq1$, and $\mu\geq1$, then
\begin{equation}\label{}
(1+x)^{\mu}\geq1+\frac{k \mu}{k+1} x+[(k+1)^{\mu}-(1+\frac{\mu}{k+1})k^{\mu}]x^{\mu}\geq 1+[(k+1)^{\mu}-k^{\mu}]x^{\mu}\geq 1+(2^{\mu}-1)x^{\mu}.
\end{equation}

Proof. If $x=0$, the inequality is trivial. Otherwise, let $f(\mu,x)=\frac{(1+x)^{\mu}-\frac{k\mu}{k+1} x-1}{x^{\mu}}$. Then, $\frac{\partial f}{\partial x}=\frac{\mu x^{\mu-1}[1+\frac{k(\mu-1)}{k+1}x-(1+x)^{\mu-1}]}{x^{2\mu}}$. When $0\leq x\leq \frac{1}{k}, k\geq1$ and $\mu\geq1$, it is easy to check that $1+\frac{k(\mu-1)}{k+1}x\leq(1+x)^{\mu-1}$. Thus, $\frac{\partial f}{\partial x}\leq0$, $f(\mu,x)$ is a decreasing function of $x$, i.e. $ f(\mu,x)\geq f(\mu,\frac{1}{k})=(k+1)^{\mu}-(1+\frac{\mu}{k+1})k^{\mu}$. Thus we have $(1+x)^{\mu}\geq1+\frac{k \mu}{k+1} x+[(k+1)^{\mu}-(1+\frac{\mu}{k+1})k^{\mu}]x^{\mu}$.

Since $kx\geq (kx)^{\mu}$, for $kx\in [0,1]$ and $\mu\geq1$, one gets $1+\frac{k \mu}{k+1} x+[(k+1)^{\mu}-(1+\frac{\mu}{k+1})k^{\mu}]x^{\mu}=1+\frac{\mu}{k+1}[kx-(kx)^{\mu}]+[(k+1)^{\mu}-k^{\mu}]x^{\mu}\geq 1+[(k+1)^{\mu}-k^{\mu}]x^{\mu}$. Let $g(\mu,k)=(k+1)^{\mu}-k^{\mu}-2^{\mu}+1$. So, $\frac{\partial g}{\partial k}=\mu[(k+1)^{\mu-1}-k^{\mu-1}]$. If $\mu\geq1$ with $k\geq1$, then $\frac{\partial g}{\partial k}\geq 0$, which implies that $g(\mu,k)$ is an increasing function of $k$, i.e. $g(\mu,k)\geq g(\mu,1)=0$, we obtain $(k+1)^{\mu}-k^{\mu}\geq 2^{\mu}-1$. Altogether, we can get $(1+x)^{\mu}\geq1+\frac{k \mu}{k+1} x+[(k+1)^{\mu}-(1+\frac{\mu}{k+1})k^{\mu}]x^{\mu}\geq1+[(k+1)^{\mu}-k^{\mu}]x^{\mu}\geq 1+(2^{\mu}-1)x^{\mu}$.

\textbf{Theorem 2}. For arbitrary tripartite quantum state $\rho_{A|B_{1}B_{2}}\in \mathcal{H}_{A}\otimes \mathcal{H}_{B_{1}}\otimes \mathcal{H}_{B_{2}}$, postulate $E^{\alpha_{c}}$ is a monogamous entanglement measure of the quantum states.

(1) If $E(\rho_{A|B_{1}})\geq \gamma E(\rho_{A|B_{2}})$, then
\begin{equation}\label{}
 \begin{aligned}
 E^{\eta}(\rho_{A|B_{1}B_{2}}) \geq
E^{\eta}(\rho_{A|B_1})+\frac{k t}{k+1}E^{\eta-\alpha_{c}}(\rho_{A|B_1})E^{\alpha_{c}}(\rho_{A|B_2})
 +[(k+1)^{t}-(1+\frac{t}{k+1})k^{t}]E^{\eta}(\rho_{A|B_2}).
\end{aligned}
\end{equation}

(2) If $\gamma E(\rho_{A|B_{1}})\leq E(\rho_{A|B_{2}})$, then
\begin{equation}\label{}
 \begin{aligned}
 E^{\eta}(\rho_{A|B_{1}B_{2}}) \geq
E^{\eta}(\rho_{A|B_2})+\frac{k t}{k+1}E^{\eta-\alpha_{c}}(\rho_{A|B_2})E^{\alpha_{c}}(\rho_{A|B_1})
 +[(k+1)^{t}-(1+\frac{t}{k+1})k^{t}]E^{\eta}(\rho_{A|B_1}),
\end{aligned}
\end{equation}
for $\eta\geq\alpha_{c}$, $\gamma\geq1$, where $t=\frac{\eta}{\alpha_{c}}$, $k=\gamma^{\alpha_{c}}$.

Proof. From the inequality (1), we can deduce
\begin{equation}\label{}
 \begin{aligned}
 E^{\eta}(\rho_{A|B_{1}B_{2}}) \geq
[E^{\alpha_{c}}(\rho_{A|B_1})+E^{\alpha_{c}}(\rho_{A|B_2})]^{t}.
\end{aligned}
\end{equation}

If $E(\rho_{A|B_{1}})\geq \gamma E(\rho_{A|B_{2}})$, according to Lemma 3, we get
\begin{equation}\label{}
 \begin{aligned}
 E^{\eta}(\rho_{A|B_{1}B_{2}}) \geq
E^{\eta}(\rho_{A|B_1})+\frac{k t}{k+1}E^{\eta-\alpha_{c}}(\rho_{A|B_1})E^{\alpha_{c}}(\rho_{A|B_2})
 +[(k+1)^{t}-(1+\frac{t}{k+1})k^{t}]E^{\eta}(\rho_{A|B_2}).
\end{aligned}
\end{equation}
When $\gamma E(\rho_{A|B_{1}})\leq E(\rho_{A|B_{2}})$, the similar proof gives the inequality (12).

Note that when $E(\rho_{A|B_{1}})\geq E(\rho_{A|B_{2}})$, $\gamma^{\alpha_{c}}=k=1$, we have
\begin{equation}\label{}
 \begin{aligned}
 E^{\eta}(\rho_{A|B_{1}B_{2}}) \geq
E^{\eta}(\rho_{A|B_1})+\frac{t}{2}E^{\eta-\alpha_{c}}(\rho_{A|B_1})E^{\alpha_{c}}(\rho_{A|B_2})
 +(2^{t}-\frac{t}{2}-1)E^{\eta}(\rho_{A|B_2}).
\end{aligned}
\end{equation}
When $E(\rho_{A|B_{1}})\leq E(\rho_{A|B_{2}})$, we can get the following inequality
\begin{equation}\label{}
 \begin{aligned}
 E^{\eta}(\rho_{A|B_{1}B_{2}}) \geq
E^{\eta}(\rho_{A|B_2})+\frac{t}{2}E^{\eta-\alpha_{c}}(\rho_{A|B_2})E^{\alpha_{c}}(\rho_{A|B_1})
 +(2^{t}-\frac{t}{2}-1)E^{\eta}(\rho_{A|B_1}).
\end{aligned}
\end{equation}

\textbf{Theorem 3.} For an $N$-party state $\rho_{A|B_{1}\cdots B_{N-1}}\in \mathcal{H}_{A}\otimes \mathcal{H}_{B_{1}}\otimes\cdots\otimes \mathcal{H}_{B_{N-1}}$,
and a monogamous entanglement measure $E^{\alpha_{c}}$, if $E(\rho_{A|B_{i}})\geq \gamma\sum\limits_{l=i+1}^{N-1} E(\rho_{A|B_{l}})$ for $i=1, 2, \cdots, m$ and $\gamma' E(\rho_{A|B_{j}})\leq\sum\limits_{l=j+1}^{N-1} E(\rho_{A|B_{l}})$ for $j=m+1, \cdots, N-2, \forall\ 1\leq m\leq N-3, N\geq4$, then
\begin{equation}\label{}
 \begin{aligned}
 E^{\eta}(\rho_{A|B_{1}\cdots B_{N-1}}) \geq &
 E^{\eta}(\rho_{A|B_{1}})+[(k+1)^{t}-k^{t}]E^{\eta}(\rho_{A|B_{2}})+\cdots
 +[(k+1)^{t}-k^{t}]^{m-1}E^{\eta}(\rho_{A|B_{m}})\\
& +[(k+1)^{t}-k^{t}]^{m}[(k'+1)^{t}-k'^{t}][E^{\eta}(\rho_{A|B_{m+1}})+\cdots+E^{\eta}(\rho_{A|B_{N-3}})]\\
& +[(k+1)^{t}-k^{t}]^{m}\left\{[(k'+1)^{t}-(1+\frac{t}{k'+1})k'^{t}]E^{\eta}(\rho_{A|B_{N-2}})\right.\\
& +\left.\frac{k' t}{k'+1}E^{\alpha_{c}}(\rho_{A|B_{N-2}})E^{\eta-\alpha_{c}}(\rho_{A|B_{N-1}})+E^{\eta}(\rho_{A|B_{N-1}})\right\}
\end{aligned}
\end{equation}
for $\eta\geq\alpha_{c}$, $\gamma\geq1$, $\gamma'\geq1$, where $t=\frac{\eta}{\alpha_{c}}$, $k=\gamma^{\alpha_{c}}$, $k'=\gamma'^{\alpha_{c}}$.

Proof. From Theorem 2, we can derive
\begin{equation}\label{}
 \begin{aligned}
  E^{\eta}(\rho_{A|B_{1}\cdots B_{N-1}}) & \geq
  E^{\eta}(\rho_{A|B_1})+\frac{k t}{k+1}E^{\eta-\alpha_{c}}(\rho_{A|B_1})\sum\limits_{l=2}^{N-1} E^{\alpha_{c}}(\rho_{A|B_{l}})\\
   & \quad+[(k+1)^{t}-(1+\frac{t}{k+1})k^{t}](\sum\limits_{l=2}^{N-1} E^{\alpha_{c}}(\rho_{A|B_{l}}))^t\\
  & \geq E^{\eta}(\rho_{A|B_{1}})+ [(k+1)^{t}-k^{t}]E^{\eta}(\rho_{A|B_{2}})+\cdots+[(k+1)^{t}-k^{t}]^{m-2}E^{\eta}(\rho_{A|B_{m-1}})\\
  & \quad +[(k+1)^{t}-k^{t}]^{m-1}\left[E^{\eta}(\rho_{A|B_{m}})+\frac{k t}{k+1}E^{\eta-\alpha_{c}}(\rho_{A|B_{m}})\sum\limits_{l=m+1}^{N-1} E^{\alpha_{c}}(\rho_{A|B_{l}})\right.\\
  & \quad+\left.[(k+1)^{t}-(1+\frac{t}{k+1})k^{t}]\left(\sum\limits_{l=m+1}^{N-1} E^{\alpha_{c}}(\rho_{A|B_{l}})\right)^{t}\right].
\end{aligned}
\end{equation}
By iterative use of inequality (11), we have the second inequality. As a matter of fact, the conditions $1+\frac{k \mu}{k+1} x+[(k+1)^{\mu}-(1+\frac{\mu}{k+1})k^{\mu}]x^{\mu}\geq 1+[(k+1)^{\mu}-k^{\mu}]x^{\mu}$ and $E(\rho_{A|B_{i}})\geq \gamma\sum\limits_{l=i+1}^{N-1} E(\rho_{A|B_{l}})$, $i=1, 2, \cdots, m$ have been used.

With a similar procedure as $\gamma' E(\rho_{A|B_{j}})\leq \sum\limits_{l=j+1}^{N-1} E(\rho_{A|B_{l}})$ for $j=m+1, \cdots, N-2$, one finds
\begin{equation}\label{}
 \begin{aligned}
\left(\sum\limits_{l=m+1}^{N-1} E^{\alpha_{c}}(\rho_{A|B_{l}})\right)^{t} & \geq
[(k'+1)^{t}-(1+\frac{t}{k'+1})k'^{t}]E^{\eta}(\rho_{A|B_{m+1}})\\
& \quad+\frac{k' t}{k'+1}E^{\alpha_{c}}(\rho_{A|B_{m+1}})\left(\sum\limits_{l=m+2}^{N-1} E^{\alpha_{c}}(\rho_{A|B_{l}})\right)^{t-1}+\left(\sum\limits_{l=m+2}^{N-1} E^{\alpha_{c}}(\rho_{A|B_{l}})\right)^{t}\\
& \geq
[(k'+1)^{t}-k'^{t}][E^{\eta}(\rho_{A|B_{m+1}})+\cdots+E^{\eta}(\rho_{A|B_{N-3}})]\\
& \quad+[(k'+1)^{t}-(1+\frac{t}{k'+1})k'^{t}]E^{\eta}(\rho_{A|B_{N-2}})+\frac{k' t}{k'+1}E^{\alpha_{c}}(\rho_{A|B_{N-2}})E^{\eta-\alpha_{c}}(\rho_{A|B_{N-1}})\\
& \quad+E^{\eta}(\rho_{A|B_{N-1}}).
\end{aligned}
\end{equation}
By using inequalities (18) and (19), we can obtain Theorem 3. In fact, we also use the condition $1+\frac{k' \mu}{k'+1} x+[(k'+1)^{\mu}-(1+\frac{\mu}{k'+1})k'^{\mu}]x^{\mu}\geq 1+[(k'+1)^{\mu}-k'^{\mu}]x^{\mu}$.

We note that if $E(\rho_{A|B_{i}})\geq \sum\limits_{l=i+1}^{N-1} E(\rho_{A|B_{l}})$ for $i=1, 2, \cdots, m$ and $E(\rho_{A|B_{j}})\leq \sum\limits_{l=j+1}^{N-1} E(\rho_{A|B_{l}})$ for $j=m+1, \cdots, N-2, \forall\ 1\leq m\leq N-3, N\geq4$, then $\gamma^{\alpha_{c}}=k=1$, $\gamma'^{\alpha_{c}}=k'=1$, we can obtain
\begin{equation}\label{}
 \begin{aligned}
 E^{\eta}(\rho_{A|B_{1}\cdots B_{N-1}}) \geq &
 E^{\eta}(\rho_{A|B_{1}})+(2^t-1)E^{\eta}(\rho_{A|B_{2}})+\cdots
 +(2^t-1)^{m-1}E^{\eta}(\rho_{A|B_{m}})\\
& +(2^t-1)^{m+1}[E^{\eta}(\rho_{A|B_{m+1}})+\cdots+E^{\eta}(\rho_{A|B_{N-3}})]\\
& +(2^t-1)^{m}\left\{(2^t-\frac{t}{2}-1)E^{\eta}(\rho_{A|B_{N-2}})\right.\\
& +\left.\frac{t}{2}E^{\alpha_{c}}(\rho_{A|B_{N-2}})E^{\eta-\alpha_{c}}(\rho_{A|B_{N-1}})+E^{\eta}(\rho_{A|B_{N-1}})\right\}.
\end{aligned}
\end{equation}

\section{tighter polygamy relations of multiparty quantum entanglement}

We use $\rho_{A|B_{1}\cdots B_{N-1}}$ denote the state of a multipartite system with a finite dimensional Hilbert space $\mathcal{H}_{A}\otimes \mathcal{H}_{B_1}\otimes\cdots\otimes\ \mathcal{H}_{B_{N-1}}$. Assume $E_{a}$ is an entanglement of assistance of the quantum states which is defined in Refs. [14, 29]. $E_{a}$ is said to be polygamy if the following inequality holds
\begin{equation}\label{}
E_{a}^{\beta_{c}}(\rho_{A|B_{1}\cdots B_{N-1}})\leq \sum\limits_{i=1}^{N-1} E_{a}^{\beta_{c}}(\rho_{A|B_{i}}),
\end{equation}
where $\rho_{A|B_{i}}=\text{tr}_{B_{1}\cdots B_{i-1}B_{i+1}\cdots B_{N-1}}(\rho_{A|B_{1}\cdots B_{N-1}})$. For the given entanglement of assistance $E_{a}$, $\beta_{c}$ is the supremum exponent for $E_{a}^{\beta_{c}}$ to be polygamy. We first prove the following conclusions.

\textbf{Lemma 4.} For $x\geq m\geq1$ and $0\leq\mu\leq1$, then
\begin{equation}\label{}
(1+x)^{\mu}\leq x^{\mu}+(m+1)^{\mu}-m^{\mu}.
\end{equation}

Proof. Let $f(\mu,x)=(1+x)^{\mu}-{x^{\mu}}$. Then, $\frac{\partial f}{\partial x}= \mu[(1+x)^{\mu-1}-x^{\mu-1}]$. When $x\geq m\geq1$ and $0\leq\mu\leq1$, it is straightforward to verify that $(1+x)^{\mu-1}\leq x^{\mu-1}$. Thus, $\frac{\partial f}{\partial x}\leq 0$, $f(\mu,x)$ is a decreasing function of $x$, i.e. $ f(\mu,x)\leq f(\mu,m)=(m+1)^{\mu}-m^{\mu}$. Thus we get $(1+x)^{\mu}\leq x^{\mu}+(m+1)^{\mu}-m^{\mu}$.

\textbf{Lemma 5.} For $a_{1}\geq a_{2}\geq \cdots \geq a_{n}\geq0$ and $0\leq\mu\leq1$, then
\begin{equation}\label{}
(a_{1}+a_{2}+\cdots+a_{n})^{\mu} \leq a_{1}^{\mu}+(2^{\mu}-1)a_{2}^{\mu}+\cdots+[n^{\mu}-(n-1)^{\mu}]a_{n}^{\mu}.
\end{equation}

Proof. We have already noted that the inequality (23) is true for the case that $n=1$. Now assume that $n=m$ the inequality (23) holds with $m>1$. Thus we have
\begin{equation}\label{}
\begin{aligned}
(a_{1}+a_{2}+\cdots+a_{m})^{\mu} \leq a_{1}^{\mu}+(2^{\mu}-1)a_{2}^{\mu}+\cdots+[m^{\mu}-(m-1)^{\mu}]a_{m}^{\mu}.
\end{aligned}
\end{equation}

Now consider the case that $n=m+1$. If $a_{m+1}=0$, the inequality is true. Otherwise, let $\tau=\frac{a_{1}+a_{2}+\cdots+a_{m}}{a_{m+1}}$, since $a_{1}\geq a_{2}\geq \cdots \geq a_{m+1}>0$, then $\tau\geq m$,

\begin{equation}\label{}
\begin{aligned}
(a_{1}+a_{2}+\cdots+a_{m}+a_{m+1})^{\mu}= & a_{m+1}^{\mu}(1+\frac{a_{1}+a_{2}+\cdots+a_{m}}{a_{m+1}})^{\mu}\\
= & a_{m+1}^{\mu}(1+\tau)^{\mu}\\
\leq & a_{m+1}^{\mu}\left[\tau^{\mu}+(m+1)^{\mu}-m^{\mu}\right]\\
= & (a_{1}+a_{2}+\cdots+a_{m})^{\mu}+[(m+1)^{\mu}-m^{\mu}]a_{m+1}^{\mu},
\end{aligned}
\end{equation}
where the inequality holds due to the inequality (22).

Combining inequalities (24) and (25) yields
\begin{equation}\label{}
\begin{aligned}
(a_{1}+a_{2}+\cdots+a_{m+1})^{\mu} \leq a_{1}^{\mu}+(2^{\mu}-1)a_{2}^{\mu}+\cdots+[(m+1)^{\mu}-m^{\mu}]a_{m+1}^{\mu}.
\end{aligned}
\end{equation}
In other words, the case that $n=m+1$ the inequality (23) is true, and the proof is completed.

\textbf{Theorem 4.} For an $N$-party state $\rho_{A|B_{1}\cdots B_{N-1}}\in \mathcal{H}_{A}\otimes \mathcal{H}_{B_{1}}\otimes\cdots\otimes \mathcal{H}_{B_{N-1}}$, suppose $E_{a}^{\beta_{c}}$ is a polygamy entanglement of assistance of the quantum states. If $E_{a}(\rho_{A|B_{i}})\geq E_{a}(\rho_{A|B_{i+1}})$ for $i=1,2,\cdots N-2$, $N\geq 3$  then
\begin{equation}\label{}
 \begin{aligned}
 E_{a}^{\eta}(\rho_{A|B_{1}\cdots B_{N-1}}) \leq
 E_{a}^{\eta}(\rho_{A|B_{1}})+(2^{t}-1)E_{a}^{\eta}(\rho_{A|B_{2}})+\cdots
 +[(N-1)^{t}-(N-2)^{t}]E_{a}^{\eta}(\rho_{A|B_{N-1}})
\end{aligned}
\end{equation}
for $0\leq\eta\leq\beta_{c}$ and $t=\frac{\eta}{\beta_{c}}$.

Proof. Without loss of generality, we may assume that, by relabeling the subsystems if necessary, the condition $E_{a}(\rho_{A|B_{i}})\geq E_{a}(\rho_{A|B_{i+1}})$ holds. From the inequality (21), we can write
\begin{equation}\label{}
 \begin{aligned}
 E_{a}^{\eta}(\rho_{A|B_{1}\cdots B_{N-1}}) \leq
[E_{a}^{\beta_{c}}(\rho_{A|B_1})+E_{a}^{\beta_{c}}(\rho_{A|B_2})+\cdots+E_{a}^{\beta_{c}}(\rho_{A|B_{N-1}})]^{t}.
\end{aligned}
\end{equation}

 If $E_{a}(\rho_{A|B_{i}})\geq E_{a}(\rho_{A|B_{i+1}})$ for $i=1,2,\cdots N-2$, from 
  Lemma 5 it follows that
\begin{equation}\label{}
 \begin{aligned}
 E_{a}^{\eta}(\rho_{A|B_{1}\cdots B_{N-1}}) \leq
 E_{a}^{\eta}(\rho_{A|B_{1}})+(2^{t}-1)E_{a}^{\eta}(\rho_{A|B_{2}})+\cdots
 +[(N-1)^{t}-(N-2)^{t}]E_{a}^{\eta}(\rho_{A|B_{N-1}}).
\end{aligned}
\end{equation}

\textbf{Remark 2}. It is easy to see that Theorem 4 is generally tighter than the polygamy relations in terms of the Hamming weight [22-25, 27].

Next, we present a mathematical result.

\textbf{Lemma 6.} For $0\leq x\leq \frac{1}{k}, k\geq1$, and $0\leq\mu\leq1$, the following inequality holds
\begin{equation}\label{}
(1+x)^{\mu}\leq 1+\frac{k^2\mu}{(k+1)^2} x+\left((k+1)^{\mu}-\left[\frac{k\mu}{(k+1)^2}+1\right]k^{\mu}\right)x^{\mu}\leq1+[(k+1)^{\mu}-k^{\mu}]x^{\mu}\leq1+(2^{\mu}-1)x^{\mu}\leq1+\mu x^{\mu}.
\end{equation}

Proof. If $x=0$, the inequality becomes trivial. Otherwise, let $f(\mu,x)=\frac{(1+x)^{\mu}-\frac{k^2\mu}{(k+1)^2} x-1}{x^{\mu}}$. Then, $\frac{\partial f}{\partial x}=\frac{\mu x^{\mu-1}[1+\frac{k^2(\mu-1)}{(k+1)^2}x-(1+x)^{\mu-1}]}{x^{2\mu}}$. When $0\leq x\leq \frac{1}{k}$,  $k\geq1$ and $1\geq\mu\geq0$, it is easy to prove that $1+\frac{k^2(\mu-1)}{(k+1)^2}x\geq(1+x)^{\mu-1}$. Thus, $\frac{\partial f}{\partial x}\geq0$, $f(\mu,x)$ is an increasing function of $x$, i.e. $ f(\mu,x)\leq f(\mu,\frac{1}{k})=(k+1)^{\mu}-[\frac{k\mu}{(k+1)^2}+1]k^{\mu}$. Then $(1+x)^{\mu}\leq1+\frac{k^2\mu}{(k+1)^2} x+\left((k+1)^{\mu}-\left[\frac{k\mu}{(k+1)^2}+1\right]k^{\mu}\right)x^{\mu}$ holds.

Due to $kx\leq (kx)^{\mu}$, for $kx\in [0,1]$ and $1\geq\mu\geq0$, we find $1+\frac{k^2\mu}{(k+1)^2} x+\left((k+1)^{\mu}-\left[\frac{k\mu}{(k+1)^2}+1\right]k^{\mu}\right)x^{\mu}=1+\frac{k\mu}{(k+1)^2}[kx-(kx)^{\mu}]+[(k+1)^{\mu}-k^{\mu}]x^{\mu}\leq1
+[(k+1)^{\mu}-k^{\mu}]x^{\mu}$. Let $g(\mu,k)=(k+1)^{\mu}-k^{\mu}-2^{\mu}+1$. So $\frac{\partial g}{\partial k}=\mu[(k+1)^{\mu-1}-k^{\mu-1}]$. For $0\leq\mu\leq1$ and $k\geq1$, we have $\frac{\partial g}{\partial k}\leq 0$, and it follows that $g(\mu,k)$ is a decreasing function of $k$, i.e. $g(\mu,k)\leq g(\mu,1)=0$. Hence, one has $(k+1)^{\mu}-k^{\mu}\leq 2^{\mu}-1$. It is clear that $\mu\geq 2^{\mu}-1$, for $1\geq\mu\geq0$. Collecting all these results we get $(1+x)^{\mu}\leq 1+\frac{k^2\mu}{(k+1)^2} x+\left((k+1)^{\mu}-\left[\frac{k\mu}{(k+1)^2}+1\right]k^{\mu}\right)x^{\mu}\leq1+[(k+1)^{\mu}-k^{\mu}]x^{\mu}\leq1+(2^{\mu}-1)x^{\mu}\leq1+\mu x^{\mu}$.

\textbf{Theorem 5.} For arbitrary tripartite quantum state $\rho_{A|B_{1}B_{2}}\in \mathcal{H}_{A}\otimes \mathcal{H}_{B_{1}}\otimes \mathcal{H}_{B_{2}}$, suppose $E_{a}^{\beta_{c}}$ is a polygamy entanglement of assistance of the quantum states.

(1) If $E_{a}(\rho_{A|B_{1}})\geq \gamma E_{a}(\rho_{A|B_{2}})$, then
\begin{equation}\label{}
 \begin{aligned}
 E_{a}^{\eta}(\rho_{A|B_{1}B_{2}}) \leq
E_{a}^{\eta}(\rho_{A|B_1})+\frac{k^2t}{(k+1)^2}E_{a}^{\eta-\beta_{c}}(\rho_{A|B_1})E_{a}^{\beta_{c}}(\rho_{A|B_2})
 +\left((k+1)^{t}-\left[\frac{kt}{(k+1)^2}+1\right]k^{t}\right)E_{a}^{\eta}(\rho_{A|B_2}).
\end{aligned}
\end{equation}

(2) If $\gamma E_{a}(\rho_{A|B_{1}})\leq E_{a}(\rho_{A|B_{2}})$, then
\begin{equation}\label{}
 \begin{aligned}
 E_{a}^{\eta}(\rho_{A|B_{1}B_{2}}) \leq
E_{a}^{\eta}(\rho_{A|B_2})+\frac{k^2t}{(k+1)^2}E_{a}^{\eta-\beta_{c}}(\rho_{A|B_2})E_{a}^{\beta_{c}}(\rho_{A|B_1})
 +\left((k+1)^{t}-\left[\frac{kt}{(k+1)^2}+1\right]k^{t}\right)E_{a}^{\eta}(\rho_{A|B_1}),
\end{aligned}
\end{equation}
for $0\leq\eta\leq\beta_{c}$, $\gamma\geq1$, where $t=\frac{\eta}{\beta_{c}}$, $k=\gamma^{\beta_{c}}$.

Proof. From the inequality (21), one can deduce that
\begin{equation}\label{}
 \begin{aligned}
 E_{a}^{\eta}(\rho_{A|B_{1}B_{2}}) \leq
[E_{a}^{\beta_{c}}(\rho_{A|B_1})+E_{a}^{\beta_{c}}(\rho_{A|B_2})]^{t}.
\end{aligned}
\end{equation}

If $E_{a}(\rho_{A|B_{1}})\geq \gamma E_{a}(\rho_{A|B_{2}})$, using Lemma 6 we find
\begin{equation}\label{}
 \begin{aligned}
 E_{a}^{\eta}(\rho_{A|B_{1}B_{2}})\leq & E_{a}^{\eta}(\rho_{A|B_1})+\frac{k^2t}{(k+1)^2}E_{a}^{\eta-\beta_{c}}(\rho_{A|B_1})E_{a}^{\beta_{c}}(\rho_{A|B_2})
 +\left((k+1)^{t}-\left[\frac{kt}{(k+1)^2}+1\right]k^{t}\right)E_{a}^{\eta}(\rho_{A|B_2}).
\end{aligned}
\end{equation}
When $\gamma E_{a}(\rho_{A|B_{1}})\leq E_{a}(\rho_{A|B_{2}})$, the inequality (32) has a similar proof.

We need to note, if $E_{a}(\rho_{A|B_{1}})\geq E_{a}(\rho_{A|B_{2}})$, $\gamma^{\beta_{c}}=k=1$, then we arrive at
\begin{equation}\label{}
 \begin{aligned}
 E_{a}^{\eta}(\rho_{A|B_{1}B_{2}}) \leq
E_{a}^{\eta}(\rho_{A|B_1})+\frac{t}{4}E_{a}^{\eta-\beta_{c}}(\rho_{A|B_1})E_{a}^{\beta_{c}}(\rho_{A|B_2})
 +(2^{t}-\frac{t}{4}-1)E_{a}^{\eta}(\rho_{A|B_2}).
\end{aligned}
\end{equation}

If $E_{a}(\rho_{A|B_{1}})\leq E_{a}(\rho_{A|B_{2}})$, then
\begin{equation}\label{}
 \begin{aligned}
 E_{a}^{\eta}(\rho_{A|B_{1}B_{2}}) \leq
E_{a}^{\eta}(\rho_{A|B_2})+\frac{t}{4}E_{a}^{\eta-\beta_{c}}(\rho_{A|B_2})E_{a}^{\beta_{c}}(\rho_{A|B_1})
 +(2^{t}-\frac{t}{4}-1)E_{a}^{\eta}(\rho_{A|B_1}).
\end{aligned}
\end{equation}

\textbf{Theorem 6.} For an $N$-party state $\rho_{A|B_{1}\cdots B_{N-1}}\in \mathcal{H}_{A}\otimes \mathcal{H}_{B_{1}}\otimes\cdots\otimes \mathcal{H}_{B_{N-1}}$,
and a polygamy entanglement of assistance $E_{a}^{\beta_{c}}$, if $E_{a}(\rho_{A|B_{i}})\geq \gamma\sum\limits_{l=i+1}^{N-1} E_{a}(\rho_{A|B_{l}})$ for $i=1,2,\cdots, m$ and $\gamma' E_{a}(\rho_{A|B_{j}})\leq \sum\limits_{l=j+1}^{N-1} E_{a}(\rho_{A|B_{l}})$ for $j=m+1,\cdots,N-2,\forall\ 1\leq m\leq N-3,N\geq4$, then
\begin{equation}\label{}
 \begin{aligned}
 E_{a}^{\eta}(\rho_{A|B_{1}\cdots B_{N-1}}) \leq &
 E_{a}^{\eta}(\rho_{A|B_{1}})+[(k+1)^{t}-k^{t}]E_{a}^{\eta}(\rho_{A|B_{2}})+\cdots
 +[(k+1)^{t}-k^{t}]^{m-1}E_{a}^{\eta}(\rho_{A|B_{m}})\\
& +[(k+1)^{t}-k^{t}]^{m}[(k'+1)^{t}-k'^{t}][E_{a}^{\eta}(\rho_{A|B_{m+1}})+\cdots+E_{a}^{\eta}(\rho_{A|B_{N-3}})]\\
& +[(k+1)^{t}-k^{t}]^{m}\left\{\left((k'+1)^{t}-\left[\frac{k't}{(k'+1)^2}+1\right]k'^{t}\right)E_{a}^{\eta}(\rho_{A|B_{N-2}})\right.\\
& +\left.\frac{k'^2t}{(k'+1)^2}E_{a}^{\beta_{c}}(\rho_{A|B_{N-2}})E_{a}^{\eta-\beta_{c}}(\rho_{A|B_{N-1}})+E_{a}^{\eta}(\rho_{A|B_{N-1}})\right\}
\end{aligned}
\end{equation}
for $0\leq\eta\leq\beta_{c}$, $\gamma\geq1$, $\gamma'\geq1$, where $t=\frac{\eta}{\beta_{c}}$, $k=\gamma^{\beta_{c}}$, $k'=\gamma'^{\beta_{c}}$.

Proof. From Theorem 5, we can deduce that
\begin{equation}\label{}
 \begin{aligned}
  E_{a}^{\eta}(\rho_{A|B_{1}\cdots B_{N-1}}) & \leq
  E_{a}^{\eta}(\rho_{A|B_1})+\frac{k^2t}{(k+1)^2}E_{a}^{\eta-\beta_{c}}(\rho_{A|B_1})\sum\limits_{l=2}^{N-1} E_{a}^{\beta_{c}}(\rho_{A|B_{l}})\\
   & \quad+\left((k+1)^{t}-\left[\frac{kt}{(k+1)^2}+1\right]k^{t}\right)(\sum\limits_{l=2}^{N-1} E_{a}^{\beta_{c}}(\rho_{A|B_{l}}))^t\\
  & \leq E_{a}^{\eta}(\rho_{A|B_{1}})+ [(k+1)^{t}-k^{t}]E_{a}^{\eta}(\rho_{A|B_{2}})+\cdots+[(k+1)^{t}-k^{t}]^{m-2}E_{a}^{\eta}(\rho_{A|B_{m-1}})\\
  & \quad +[(k+1)^{t}-k^{t}]^{m-1}\left[E_{a}^{\eta}(\rho_{A|B_{m}})+\frac{k^2t}{(k+1)^2}E_{a}^{\eta-\beta_{c}}(\rho_{A|B_{m}})\sum\limits_{l=m+1}^{N-1} E_{a}^{\beta_{c}}(\rho_{A|B_{l}})\right.\\
  & \quad+\left.\left((k+1)^{t}-\left[\frac{kt}{(k+1)^2}+1\right]k^{t}\right)\left(\sum\limits_{l=m+1}^{N-1} E_{a}^{\beta_{c}}(\rho_{A|B_{l}})\right)^{t}\right].
\end{aligned}
\end{equation}
Iterative use of inequality (31), we can get the second inequality. Here we are using the fact that $1+\frac{k^2\mu}{(k+1)^2} x+\left((k+1)^{\mu}-\left[\frac{k\mu}{(k+1)^2}+1\right]k^{\mu}\right)x^{\mu}\leq1+[(k+1)^{\mu}-k^{\mu}]x^{\mu}$ and $E_{a}(\rho_{A|B_{i}})\geq \gamma\sum\limits_{l=i+1}^{N-1} E_{a}(\rho_{A|B_{l}})$, $i=1,2,\cdots,m$.

Following a similar procedure as $\gamma' E_{a}(\rho_{A|B_{j}})\leq \sum\limits_{l=j+1}^{N-1} E_{a}(\rho_{A|B_{l}})$ for $j=m+1,\cdots,N-2$, we have
\begin{equation}\label{}
 \begin{aligned}
\left(\sum\limits_{l=m+1}^{N-1} E_{a}^{\beta_{c}}(\rho_{A|B_{l}})\right)^{t} & \leq
\left((k'+1)^{t}-\left[\frac{k't}{(k'+1)^2}+1\right]k'^{t}\right)E_{a}^{\eta}(\rho_{A|B_{m+1}})\\
& \quad+\frac{k'^2t}{(k'+1)^2}E_{a}^{\beta_{c}}(\rho_{A|B_{m+1}})\left(\sum\limits_{l=m+2}^{N-1} E_{a}^{\beta_{c}}(\rho_{A|B_{l}})\right)^{t-1}+\left(\sum\limits_{l=m+2}^{N-1} E_{a}^{\beta_{c}}(\rho_{A|B_{l}})\right)^{t}\\
& \leq
[(k'+1)^{t}-k'^{t}][E_{a}^{\eta}(\rho_{A|B_{m+1}})+\cdots+E_{a}^{\eta}(\rho_{A|B_{N-3}})]\\
& \quad+\left((k'+1)^{t}-\left[\frac{k't}{(k'+1)^2}+1\right]k'^{t}\right)E_{a}^{\eta}(\rho_{A|B_{N-2}})\\
& \quad+\frac{k'^2t}{(k'+1)^2}E_{a}^{\beta_{c}}(\rho_{A|B_{N-2}})E_{a}^{\eta-\beta_{c}}(\rho_{A|B_{N-1}})+E_{a}^{\eta}(\rho_{A|B_{N-1}}).
\end{aligned}
\end{equation}
Theorem 6 can be obtained by combining inequalities (38) with (39). We also use the fact that $1+\frac{k'^2\mu}{(k'+1)^2} x+\left((k'+1)^{\mu}-\left[\frac{k'\mu}{(k'+1)^2}+1\right]k'^{\mu}\right)x^{\mu}\leq1+[(k'+1)^{\mu}-k'^{\mu}]x^{\mu}$.

Let us note that if $E_{a}(\rho_{A|B_{i}})\geq \sum\limits_{l=i+1}^{N-1} E_{a}(\rho_{A|B_{l}})$ for $i=1, 2, \cdots, m$ and $E_{a}(\rho_{A|B_{j}})\leq \sum\limits_{l=j+1}^{N-1} E_{a}(\rho_{A|B_{l}})$ for $j=m+1, \cdots, N-2, \forall\ 1\leq m\leq N-3, N\geq4$, then $\gamma^{\beta_{c}}=k=1$, $\gamma'^{\beta_{c}}=k'=1$, it follows that
\begin{equation}\label{}
 \begin{aligned}
 E_{a}^{\eta}(\rho_{A|B_{1}\cdots B_{N-1}}) \leq &
 E_{a}^{\eta}(\rho_{A|B_{1}})+(2^t-1)E_{a}^{\eta}(\rho_{A|B_{2}})+\cdots
 +(2^t-1)^{m-1}E_{a}^{\eta}(\rho_{A|B_{m}})\\
& +(2^t-1)^{m+1}[E_{a}^{\eta}(\rho_{A|B_{m+1}})+\cdots+E_{a}^{\eta}(\rho_{A|B_{N-3}})]\\
& +(2^t-1)^{m}\left\{(2^t-\frac{t}{4}-1)E_{a}^{\eta}(\rho_{A|B_{N-2}})\right.\\
& +\left.\frac{t}{4}E_{a}^{\beta_{c}}(\rho_{A|B_{N-2}})E_{a}^{\eta-\beta_{c}}(\rho_{A|B_{N-1}})+E_{a}^{\eta}(\rho_{A|B_{N-1}})\right\}.
\end{aligned}
\end{equation}

To see the tightness of our inequalities, we give some examples below. We using the concurrence as a bipartite measure of entanglement, the concurrence of assistance as a bipartite entanglement of assistance.

\emph{Example 1}: For the four-qubit W state [27]
\begin{equation}\label{}
|\psi\rangle_{ABCD}=\frac{1}{2}(|1000\rangle+|0100\rangle+|0010\rangle+|0001\rangle),
\end{equation}
one finds that the concurrence $C(|\varphi\rangle_{A|BCD})=\frac{\sqrt{3}}{2}$, $C(\rho_{A|B})=C(\rho_{A|C})=C(\rho_{A|D})=\frac{1}{2}$ and the concurrence of assistance $C_{a}(|\varphi\rangle_{A|BCD})=\frac{\sqrt{3}}{2}$, $C_{a}(\rho_{A|B})=C_{a}(\rho_{A|C})=C_{a}(\rho_{A|D})=\frac{1}{2}$. The state (41) saturates the inequality (7) and (29), but the inequality in terms of the Hamming weight cannot reach the bound  $C(|\varphi\rangle_{A|BCD})=C_{a}(|\varphi\rangle_{A|BCD})=\frac{\sqrt{3}}{2}$. Thus Theorems 1 and 4 are better than the monogamy and polygamy relations in terms of the Hamming weight [22-25, 27].

\emph{Example 2}: Under local unitary operations, the three-qubit pure state can be written as [30]
\begin{equation}\label{}
|\varphi\rangle_{ABC}=\lambda_{0}|000\rangle+\lambda_{1}\text{e}^{\iota\phi}|100\rangle+\lambda_{2}|101\rangle+\lambda_{3}|110\rangle+\lambda_{4}|111\rangle,
\end{equation}
where $\iota=\sqrt{-1}$, $0\leq\phi\leq\pi$, $\lambda_{s}\geq0$, $s=0, 1, 2, 3, 4$, and $\sum\lambda^2_{s}=1$.
Set $\lambda_{0}=\frac{1}{2}$, $\lambda_{1}=\lambda_{4}=\frac{\sqrt{2}}{12}$, $\lambda_{2}=\frac{\sqrt{2}}{2}$, $\lambda_{3}=\frac{\sqrt{2}}{3}$.

After some analysis of the concurrence, we can get $C(\rho_{A|B})=\frac{\sqrt{2}}{3}, C(\rho_{A|C})=\frac{\sqrt{2}}{2}$, $C(|\varphi\rangle_{A|BC})=C_{a}(|\varphi\rangle_{A|BC})=\frac{\sqrt{106}}{12}$. One can explicitly see that our lower bound is larger than the results in [20, 21, 26, 27], as illustrated in Fig.1.

\begin{figure}
\begin{center}
{\includegraphics[scale=0.275]{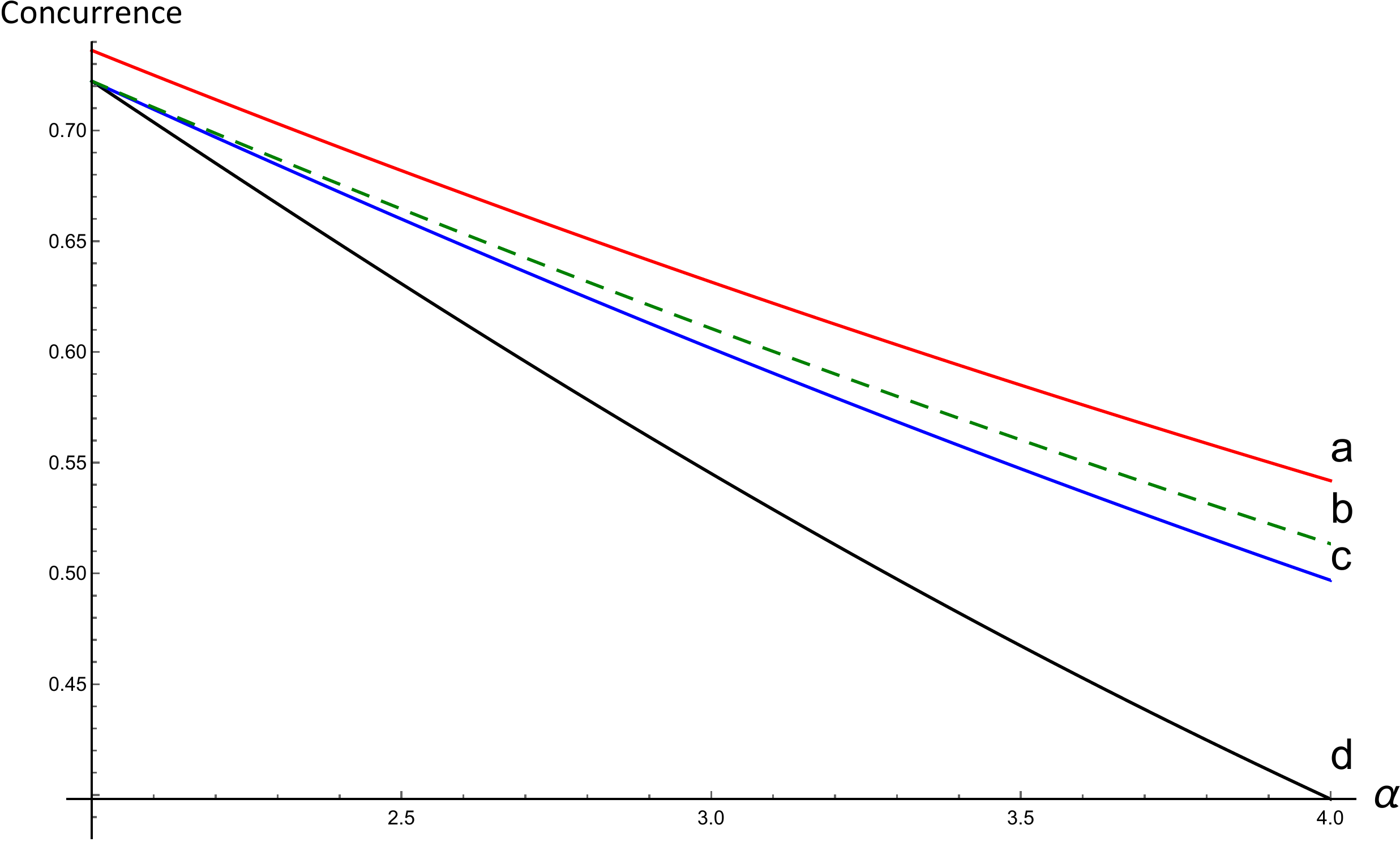}} \caption[Illustration of
]{(Color online) The ( red solid ) line
$a$ represents the $C^\alpha(|\varphi\rangle_{A|BC})$ in Example 2. The (green dashed ) line
$b$ represents the lower bound given by inequality (11) with $k=2$. The ( blue ) line $c$ represents the lower bound from the result in [26] with $k=2$. The ( black ) line $d$ represents the lower bound from the result in [20, 21, 27].}
\end{center}
\end{figure}

Straightforward calculation of the concurrence of assistance, we have  the $C_{a}(\rho_{A|B})=\frac{\sqrt{34}}{12}, C_{a}(\rho_{A|C})=\frac{\sqrt{74}}{12}$. One can explicitly see that our upper bound is smaller than the results in [22-27], as shown in Fig.2.

\begin{figure}
\begin{center}
{\includegraphics[scale=0.225]{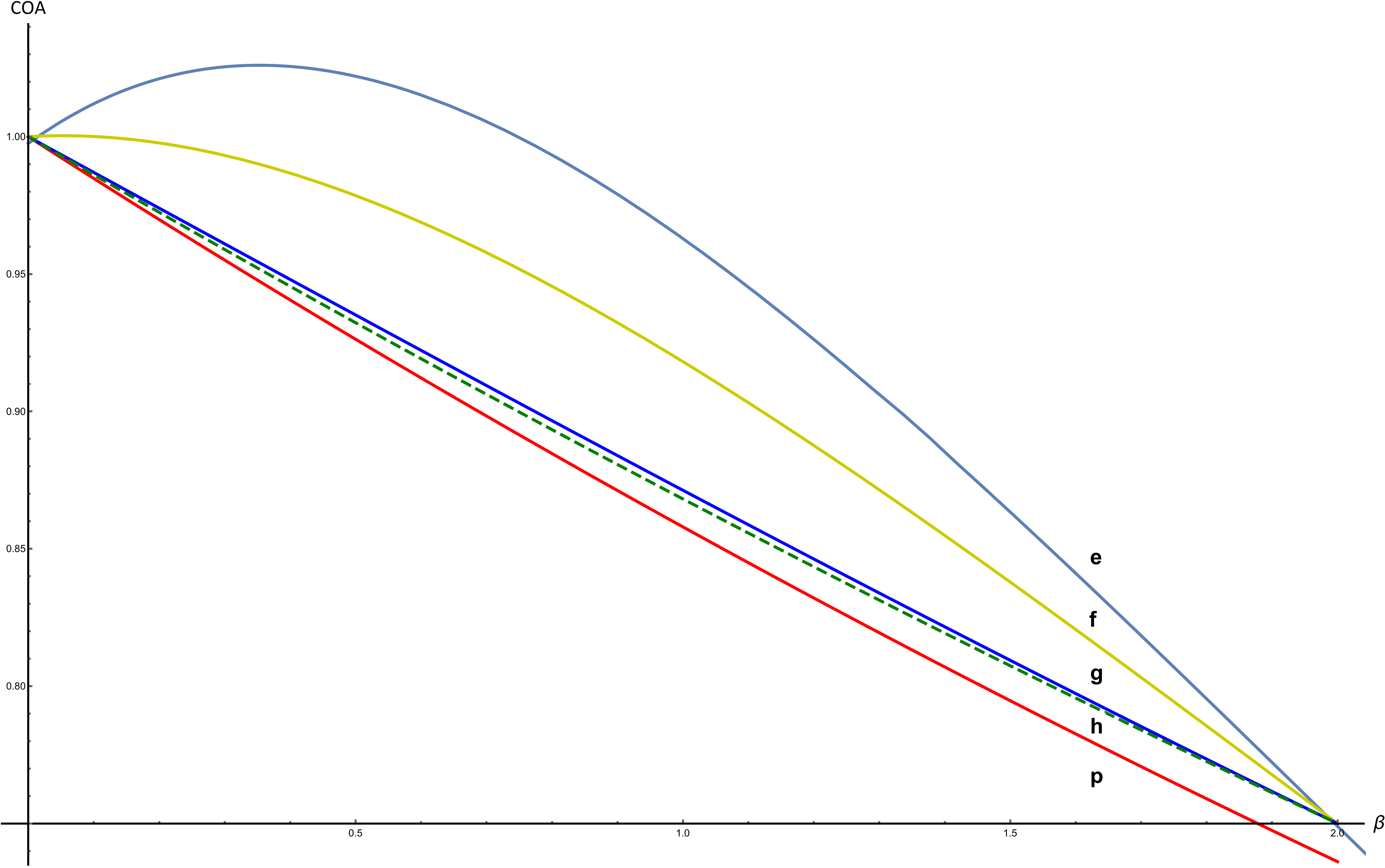}} \caption[Illustration of
]{(Color online) The ( red solid ) line
$p$ represents the $C_{a}^\beta(|\varphi\rangle_{A|BC})$ in Example 2. The (green dashed ) line
$h$ represents the upper bound given by inequality (31) with $k=2$. The ( blue ) line $g$ represents the upper bound from the result in [25, 26] with $k=2$. The ( yellow ) line $f$ represents the upper bound from the result in [27]. The ( black ) line $e$ represents the upper bound from the result in [22-24].}
\end{center}
\end{figure}

\section{CONCLUSION}

Multipartite entanglement can be regarded as a fundamental problem in the theory of quantum entanglement. It has attracted increasing interest over the last 20 years.  Our results may contribute to a fuller understanding of the multiparty quantum entanglement. By using the power of the bipartite measure of entanglement and the entanglement of assistance, we have proposed a new class of tight monogamy and polygamy relations of multiparty entanglement for arbitrary quantum states. We show that these new monogamy relations of multiparty entanglement with larger lower bounds than the existing monogamy relations [20, 21, 26, 27], for $\eta\geq\alpha_{c}$. For $0\leq\eta\leq\beta_{c}$, these new polygamy relations of multiparty entanglement with smaller upper bounds than the existing polygamy relations [22-27].

\vspace{0.6cm}
\acknowledgments
This work was supported by the National Natural Science Foundation of China under Grant No: 11475054, the Hebei Natural Science Foundation of China under Grant No: A2018205125.

\end{document}